\documentclass[sigconf,natbib=true,nonacm]{acmart}
\usepackage{graphicx}
\usepackage{array}
\usepackage{float}
\usepackage{xurl}
\usepackage{hyperref} 

\hypersetup{
    colorlinks=true,
    linkcolor=blue,
    filecolor=blue,      
    urlcolor=blue,
    pdftitle={Overleaf Example},
    pdfpagemode=FullScreen,
    }

\usepackage{wrapfig,lipsum,booktabs}

\AtBeginDocument{%
  \providecommand\BibTeX{{%
    \normalfont B\kern-0.5em{\scshape i\kern-0.25em b}\kern-0.8em\TeX}}}

  \acmConference[NA]{NA}{NA,
  2024}{NA}

\begin{document}

\title{Academic Article Recommendation Using Multiple Perspectives}

\author{Kenneth Church}
\affiliation{%
  \institution{Northeastern University}
  \city{San Jose}
  \country{CA}}
\email{k.church@northeastern.edu}

\author{Omar Alonso}
\authornote{Work does not relate to the author's position at Amazon.}
\affiliation{%
  \institution{Amazon}
  \city{Palo Alto}
  \country{CA}}
\email{omralon@amazon.com}

\author{Peter Vickers}
\affiliation{%
  \institution{Northeastern University}
  \city{Portland}
  \country{ME}}
\email{p.vickers@northeastern.edu}

\author{Jiameng Sun}
\affiliation{%
  \institution{Northeastern University}
  \city{San Jose}
  \country{CA}}
\email{sun.jiam@northeastern.edu}

\author{Abteen	Ebrahimi}
\affiliation{%
  \institution{University of Colorado}
  \city{Boulder}
  \country{CO}}
\email{Abteen.Ebrahimi@colorado.edu}

\author{Raman	Chandrasekar}
\affiliation{%
  \institution{Northeastern University}
  \city{Seattle}
  \country{WA}}
\email{r.chandrasekar@northeastern.edu}

\begin{abstract}

We argue that Content-based filtering (CBF) and Graph-based methods (GB) complement one another in Academic Search recommendations. The scientific literature can be viewed as a conversation between authors and the audience. CBF uses abstracts to infer authors' positions, and GB uses citations to infer responses from the audience. In this paper, we describe nine differences between CBF and GB, as well as synergistic opportunities for hybrid combinations. Two embeddings will be used to illustrate these opportunities: (1) Specter, a CBF method based on BERT-like deepnet encodings of abstracts, and (2) ProNE, a GB method based on spectral clustering of more than 200M papers and 2B citations from Semantic Scholar.

\end{abstract}
\maketitle
\pagestyle{plain}

\section{Introduction}

Academic Search is an information retrieval vertical widely used for scholarly search. Academic Search has other practical applications beyond recommending papers to read. For example, authors need to find papers they should cite, and program committees
and funding agencies need to assign submissions to reviewers that are well-informed and sympathetic to
the topic area. Better recommendations will improve reviews, publication quality, and move a field into new directions. Academic Search data sets are large enough for research purposes. Due to its scale, there are opportunities to transfer improvements on graph learning from academic search to recommendation
systems based on more sensitive data such as eCommerce product search, finance and traffic analysis for military intelligence.

Besides standard information needs like performing a paper or author search and looking up for citations, there is room for specific recommendations as follows:
\begin{itemize}
    \item Papers-like-this. Given a paper-id as input, recommend papers near the paper-id and return a list of papers.
\item Authors-like-this. Given a paper-id as input, recommend authors near the paper-id and return a list of authors.
\end{itemize}

Recent surveys on recommender systems in academic search \cite{bai2019scientific,KreutzS22} group the literature into four approaches:

\begin{enumerate}
  \setlength{\itemsep}{0pt}
  \setlength{\parskip}{0pt}
 \setlength{\parsep}{0pt}
 \item Content-Based Filtering (CBF),
 \item Graph-Based methods (GB), 
 \item Collaborative Filtering (CF), and
 \item Better Together: hybrids/ensembles of above.
\end{enumerate}

This paper will focus on CBF and GB because of the availability of data
and the lack of sensitivity.\footnote{This work was inspired by insights gained from \textcolor{blue}{\href{https://jsalt2023.univ-lemans.fr/en/better-together-text-context.html}{JSALT 2023}}, which was supported by Johns Hopkins University and H2020-MSCA ESPERANTO.} 
Semantic Scholar (S2) \cite{Wade2022TheSS,Kinney2023TheSS} provides data
on more than 200M academic papers and 2B citations.
S2 supports bulk downloads as well as API access by ids such as:
MAG \cite{wang2020microsoft}, DOI, PubMed, DBLP, arXiv and ACL.\footnote{\url{https://aclanthology.org/}}
From an id, one can (often) retrieve a number of useful fields including titles, authors, abstracts,
references, citations, Specter embeddings \cite{cohan-etal-2020-specter} and more.\footnote{\label{footnote:S2_api}\url{https://api.semanticscholar.org/api-docs/}}

We have found ensembles of CBF and GB to be
very effective. Ensembles of CBF, GB and CF would be even more effective, but CF requires access to (sensitive) behavioral signals \cite{Beel2018RARDIT}.

\begin{wrapfigure}{r}{0.18\textwidth}
     \centering
    \includegraphics[width=0.19\textwidth]{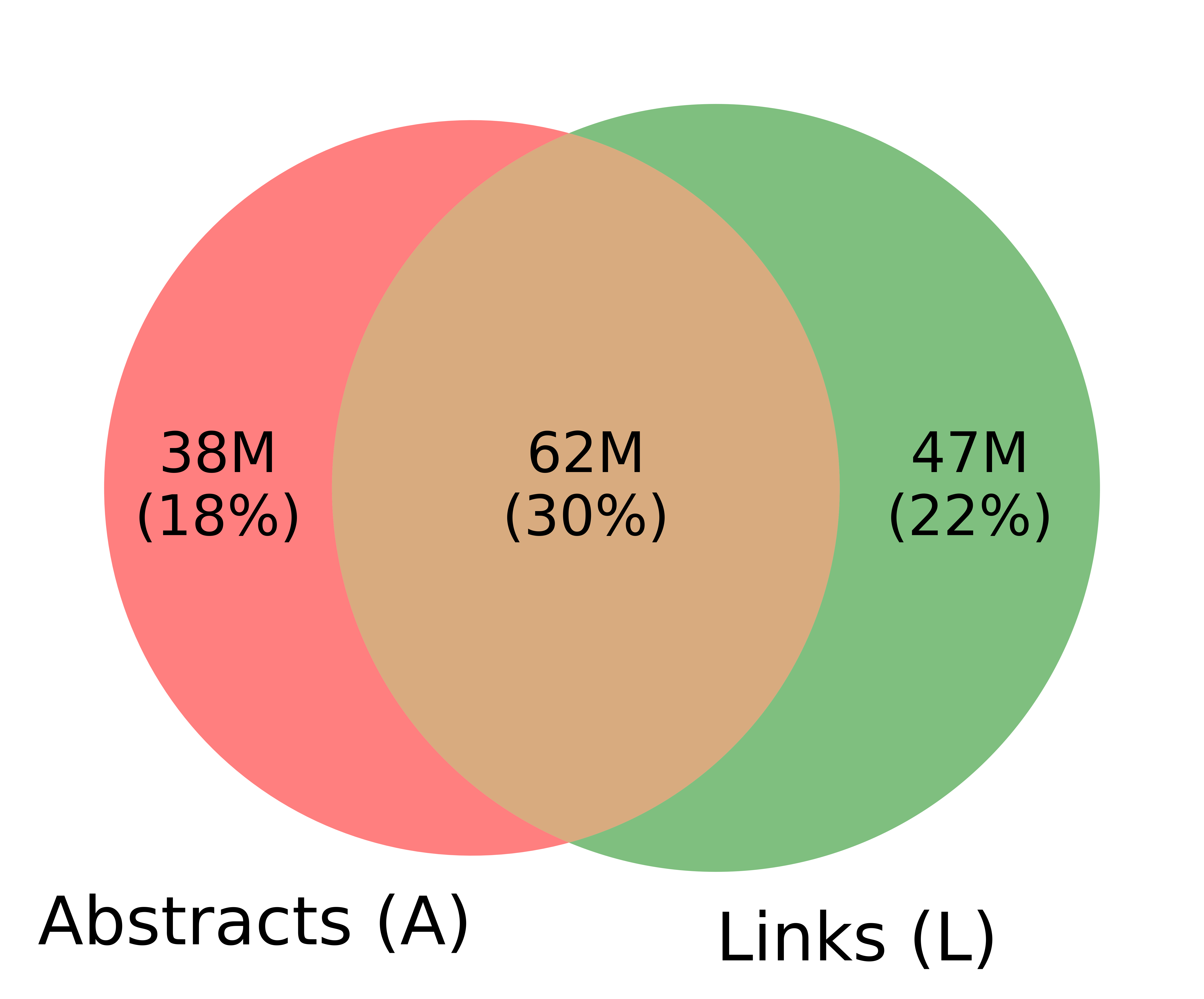}
    \caption{$|A \cap L| \approx 30\%$}
    \label{fig:venn}
\end{wrapfigure}

CBF methods in this paper focus on titles and abstracts in S2;
GB methods in this paper are based on a citation graph constructed
from citations in S2.
Some graph learning methods such as GNNs (graph neural networks) \cite{Wu2020GraphNN}
assume that most papers have both abstracts and links in the graph.
\autoref{fig:venn} shows that S2 is not like that; only 30\% of the papers in S2 have both abstracts and links,
and consequently, ensembles of CBF (titles \& abstracts) and GB (links in citation graph) have better
coverage than either method by itself.

\subsection{Scenario: Recommending Related Work}

One of the motivations for papers-like-this is to help authors write surveys and related work sections.  Papers-like-this can be helpful for the first of four subtasks:
\begin{enumerate}
  \setlength{\itemsep}{0pt}
  \setlength{\parskip}{0pt}
 \setlength{\parsep}{0pt}
    \item \textbf{candidate list generation}: list papers to discuss,
    \item \textbf{organization}: organize list by topic, time, etc.,
    \item \textbf{summarization}: summarize papers, and 
    \item \textbf{connecting the dots}: explain how papers are relevant to the present discussion
\end{enumerate}

\begin{table}[b]
\centering
\begin{tabular}{ l l l l}
\toprule 
\textbf{Task: Find papers on ...} & \textbf{Query}  & \textbf{CBF} & \textbf{GB} \\ \hline

Recommender systems &
\cite{Beel2015ResearchpaperRS} & 
\cite{Maake2019InformationPI,Beel2015ACO,Beel2013ResearchPR,Beel2013ACA,Singh2021RecommendationS} &
\cite{Nascimento2011ASI,Sugiyama2010ScholarlyPR,Sun2014LeveragingCA,Huang2019BayesianRS,Bai2017ExploitingSH} 

\\ \hline

Who should review what? &
\cite{Mimno2007ExpertiseMF} & 
\cite{Zhang2023WhySI,Anjum2022SubmissionAwareRP,Balog2008AssociatingPA,Mangaravite2016OnID,Anjum2019PaReAP,Tu2010CitationAT} & \cite{Campos2021LDAbasedTP,Jin2017IntegratingTT,Yin2013AUM,Zhang2011TopicAF,Daud2012UsingTT} \\ \hline

Citation Recommendation &
\cite{Färber2020CitationRA} &
\cite{Ali2021AnOA,Jia2017AnAO,Jebari2023ContextawareCR,Medic2021ASO,Ma2020ARO} &
\cite{Khadka2018UsingCT,Jia2018LocalIG,Chen2020ANC,Etemadi2021CollaborativeED}
 \\ \hline

RAG &
\cite{Lewis2020RetrievalAugmentedGF} &
\cite{Tan2022TegTokAT,Yu2022RetrievalaugmentedGA,Asai2021EvidentialityguidedGF,Sun2022RecitationAugmentedLM,Ivison2022HyperdecodersID}
&
\cite{Lewis2020PretrainingVP,Guu2020REALMRL,Lazaridou2022InternetaugmentedLM,Alberti2019SyntheticQC,Tay2022UnifyingLL} \\ \hline

\hline 
\end{tabular}
\caption{\label{tab:related_work}  Complementary Recommendations.}
\end{table}

\noindent
\autoref{tab:related_work} shows two lists of papers to discuss,
generated by two recommendation methods: CBF (S2 API)\footnote{\label{footnote:S2_recommendations}\url{https://api.semanticscholar.org/recommendations/v1/papers/forpaper/21321bad706a9f9dbb502588b0bb393cf15fa052?from=all-cs&fields=title,externalIds,citationCount}} and GB (ProNE \cite{zhang2019prone}).  Both  methods take as input a query (an id in S2, e.g., the second column) and output recommendations (more ids in S2).  The two methods are complementary; they return different papers (last two columns) that are relevant for different reasons.

This paper will not discuss the remaining steps mentioned above: organization, summarization and connecting the dots.  We have experimented
with RAG (Retrieval-Augmented Generation) and chatbots to address some of those steps, and also to explain why a candidate recommendation is relevant to a query.  In addition
to obvious ethical concerns, we have found chatbot summaries and explanations to be impressive
on first impression, but repetitive and uninspired on further reflection.

\section{Similarities: Embeddings}

It is standard practice to implement both CBF and GB as embeddings,
where an embedding is a matrix, $M \in \mathbb{R}^{n \times d}$.
The number of documents, $n$, is typically ${\sim}10^8$
and the number of hidden dimensions, $d$, is typically ${\sim}10^2$ (768 for BERT-like models).
The similarity of two papers, $i$ and $j$, is the cosine of their two vectors: $cos(M[i,:], M[j,;])$.   We use approximate
nearest neighbors (ANN) 
\cite{indyk1998approximate} 
to find recommendations near an input query.
This much holds for both CBF and GB,
but there are large differences between CBF and GB, as will be
discussed in the next section.\footnote{
Access to 
CBF and GB embeddings is provided on website (anonymized for blind review).
The CBF embeddings 
were supplied by S2.  They encode $n=127$ million abstracts with $d=768$ hidden dimension,
using Specter \cite{cohan-etal-2020-specter}, a BERT-like \cite{devlin-etal-2019-bert} deep net.
Note that S2 distributes more Specter vectors than abstracts because they are not allowed
to distribute many of their abstracts (personal communication).
The GB embeddings on this website encode $n=119$ million documents with $d=280$ hidden dimensions, 
using a spectral clustering \cite{Luxburg2007ATO} method called ProNE \cite{zhang2019prone}.}

\section{Complementary Perspectives}

As suggested above, CBF and GB complement one another.
We view the literature as a conversation
between authors and the audience.  CBF uses abstracts to infer authors' positions,
and GB uses citations to infer responses from the audience.

\subsection{Differences / Synergistic Opportunities}

The next nine subsubsections will discuss nine differences:

\begin{enumerate}
  \setlength{\itemsep}{0pt}
  \setlength{\parskip}{0pt}
 \setlength{\parsep}{0pt}
 
 \item \textbf{Inputs}: Titles and abstracts for CBF; 
 citations for GB.
 
  \item \textbf{Interpretations}: For CBF, large cosines indicate similar abstracts; for GB, large cosines indicate similarity in terms of random walks on citation graph.

 \item \textbf{History}: Deep networks evolved out of use cases in Natural Language Processing (NLP) whereas spectral clustering \cite{Luxburg2007ATO} has roots in Linear Algebra and was inspired, at least in part, by use cases such as traffic analysis in Applied Math.

  \item \textbf{Implementation Details}: We use Specter (a deep network) for CBF
 and ProNE (spectral clustering) for GB.
 
 \item \textbf{Computational Bottlenecks}: Deep networks are limited by computational cycles, whereas spectral clustering is limited by memory.  We use GPUs for deep networks, and terabytes of RAM to compute SVDs for spectral clustering.
 
 \item \textbf{Scale}: Larger graphs favor GB because of network effects.
 
 \item \textbf{Time Invariance}: CBF embeddings are time invariant because
 abstracts do not change after publication; GB embeddings improve as papers accumulate citations over time.
 
 \item \textbf{Priors}: GB recommendations have more citations, but are less recent (because it takes time to accumulate citations).

 \item \textbf{Corner Cases and Missing Values}: Multiple perspectives create opportunities to improve robustness and coverage with error detection and imputing missing values.
 \end{enumerate}

\subsubsection{Inputs}
As mentioned above, CBF is based on titles \& abstracts, whereas GB  is based on citations.  Many of the differences mentioned below
are consequences of these different inputs.  Abstracts are
more representative of the authors' perspectives;
citations are more representative of responses from the audience.

\subsubsection{Interpretations}
Large cosines suggest that papers are similar to one another,
but for different reasons.
For CBF methods, a large cosine implies the two abstracts use similar words in terms of
large language models (LLMs).  In contrast, for GB methods, a large cosine implies the two papers
are near one another in terms of random walks on the citation graph.

\begin{wrapfigure}{r}{0.24\textwidth}
    \centering
    \includegraphics[width=0.25\textwidth]{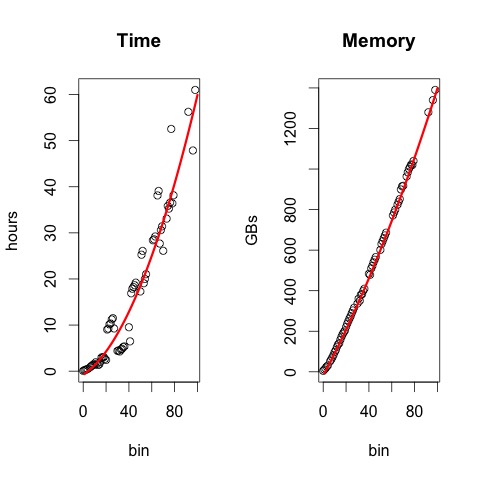}
    \caption{\label{fig:prefactor} ProNE prefactorization (SVD) takes a few days and a few TBs on larger graphs.}
\end{wrapfigure}

\subsubsection{History}

CBF and GB come from different disciplines.
It is popular these days to use Deep Networks and
LLMs from Computer Science for CBF.
For GB, it is standard practice to use Eigenvector methods
such as Page Rank \cite{page1999pagerank} from Linear Algebra.
Spectral Clustering (ProNE) is like Page Rank, but with more
hidden dimensions.  

Much of the GB literature was motivated by the traffic analysis use case, where we are given the graph of who is communicating with whom, but
we do not know what they are saying because it is assumed
that payloads are encrypted.
The literature on LLMs historically evolved from use cases in NLP where the situation is reversed.
In many NLP use cases, we are given the contents of the documents,
but not the larger context of how
documents are connected to one another.

\subsubsection{Implementation Details}

We have been using Specter and ProNE as representative instances of CBF and GB approaches, respectively, though there are many variations:

\begin{enumerate}
  \setlength{\itemsep}{0pt}
  \setlength{\parskip}{0pt}
 \setlength{\parsep}{0pt}
 \item CBF: Specter \cite{cohan-etal-2020-specter}, 
 SciNCL \cite{ostendorff-etal-2022-neighborhood},
 Sentence Encoders \cite{Hassan2019BERTEU}

 \item GB: ProNE \cite{zhang2019prone}, GraRep \cite{Cao2015GraRepLG}, Node2Vec \cite{Grover2016node2vecSF}, DeepWalk \cite{Perozzi2014DeepWalkOL}, 
 LINE \cite{Tang2015LINELI} 
 and more in surveys \cite{Cui2017ASO,Qiu2017NetworkEA}

\end{enumerate}=

\noindent
At query time, all of these embeddings use ANN (approximate nearest neighbors) to take input queries (ids in S2) and output recommendations (more ids in S2).
The differences between CBF and GB mostly involve the creation of embeddings, as well as inference and fine-tuning of deep network models.
For ProNE, we use the nodevectors package\footnote{\url{https://pypi.org/project/nodevectors/}} to create an embedding based on a citation graph from S2.
For Specter, we did not need to create vectors because they can be downloaded from S2 (footnote \ref{footnote:S2_api}).
In addition, S2 also provides a recommendation API (footnote \ref{footnote:S2_recommendations})
which is based on the FAISS\footnote{\url{https://ai.meta.com/tools/faiss/}} implementation of ANN.  Unfortunately, the recommendation API is limited to papers in Computer Science (about 10\% of their collection), because of cost constraints (personal communication).

At inference time, Specter takes a text string (typically titles and abstracts) as input,
and outputs a vector of 768 floats. 
As suggested above, since Specter vectors can be downloaded, it is not necessary
to run the model at inference time or query time.  However,
the model is available,\footnote{\url{https://huggingface.co/allenai/specter2}} if one wants to generate recommendations for an unpublished paper.  

It is also not necessary to fine-tune Specter,
but if one wanted to do that, \cite{cohan-etal-2020-specter} explains how they did it.
Citations are not used at inference time, but they
are used for fine-tuning.
Fine-tuning starts with SciBERT \cite{beltagy2019scibert}.
Specter is trained on 684K triples from S2:
$<query, pos, neg>$,
where $pos$ is a positive paper (1-2 hops from the query paper) and $neg$ is a random negative paper. 
Specter2 uses the same architecture, but is trained on 6M triples \cite{Singh2022SciRepEvalAM}.  ProNE does not use content (abstracts), but it is based on billions of citations, considerably more than a few million triples.

\subsubsection{Computational Bottlenecks}

The computational bottlenecks are quite different, as well.  
GPUs are effective for BERT-like deep nets (Specter), but 
    spectral clustering is more constrained by memory than computational cycles, and consequently,
    CPUs with TBs of RAM are preferable for computing ProNE embeddings.
The bottleneck for ProNE is an SVD\footnote{Nodevectors computes the SVD with randomized\_svd from sklearn.} on the input citation graph.
\autoref{fig:prefactor} shows time and space for the SVD
as a function of graph size (bin).
It takes a few days and a few TB of RAM for ProNE to embed the citation graph for 100 bins (200 million papers).
Availability of GPUs and TBs are enabling advances in both CBF and GB.

\begin{figure}
\centering
\includegraphics[width=0.75\linewidth]{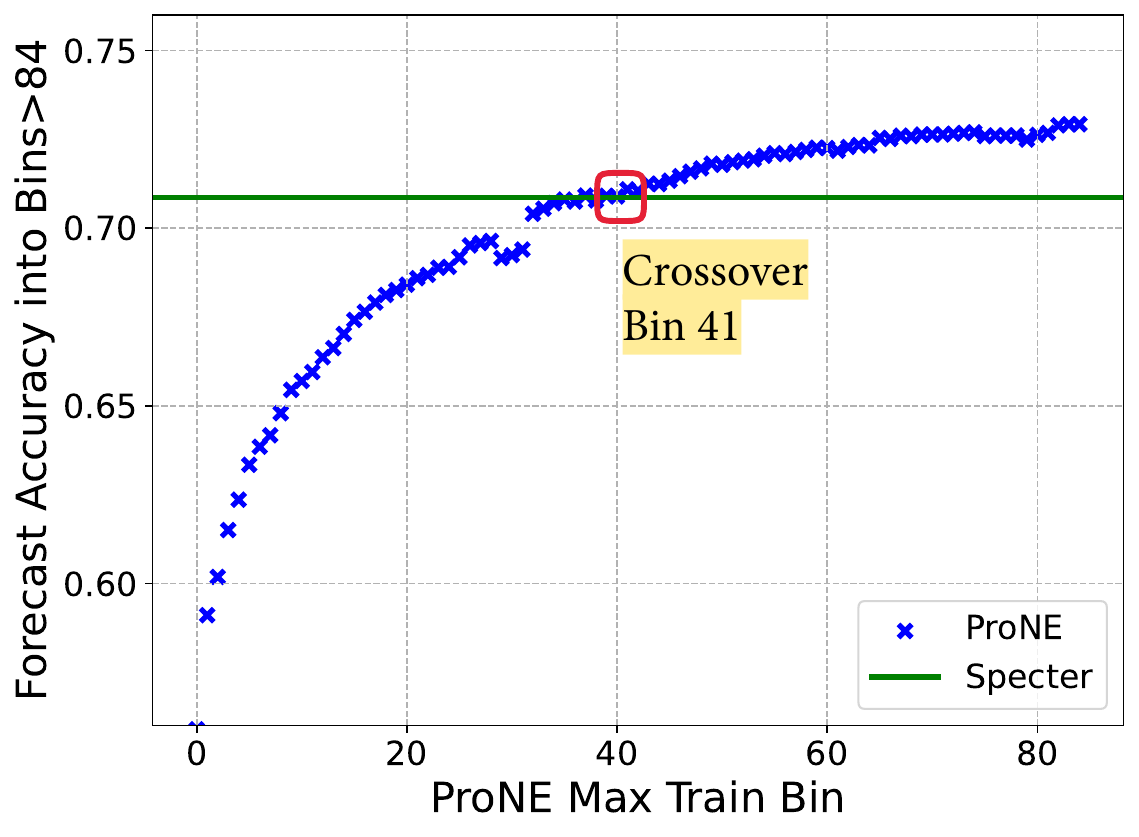}
\caption{\label{fig:prone-specter-crossover2} Larger graphs favor ProNE (GB).}
\end{figure}

\begin{table}
{\centering

\begin{tabular}{r | c c}
\toprule
 & \textbf{Vertices (Papers)} & \textbf{Edges (Links)} \\ \hline
\textbf{ ogbl-citation2} & 3M & 31M \\
 \textbf{crossover (bin 41)} & 42M & 499M \\ \bottomrule
\end{tabular}}
\caption{\label{tab:crossover} $|$OGB$| \ll$ crossover}
\end{table}

\subsubsection{Scale}

\autoref{fig:prone-specter-crossover2} is based on related work on
a citation prediction task: predict whether paper $a$ cites paper $b$.  The test set\footnote{The test set is posted at website (anonymized for blind review).} contains pairs of papers that are relatively close in the citation graph: separated by 1-4 hops.  The task
is to distinguish the 1-hop pairs from the rest.
Performance is shown for a Specter model trained on bin 85
and 100 ProNE models trained on 100 subgraphs of increasing size
(number of bins for training).
The evaluation on bin 84 shows larger graphs favor ProNE (GB).\footnote{Another evaluation shows performance not only improves with $t$ (number of bins for training) but also degrades with $h$ (forecasting horizon in terms of bins after $t$).}

GNNs attempt to combine abstracts and citations into a single
model that can be applied at inference time to novel inputs.
However, \autoref{fig:prone-specter-crossover2} suggests
that the optimal combination of nodes and edges varies by the size of the graph.  
Much of the prior work is based on benchmarks such as OGB (Open Graph Benchmark) \cite{Hu2021OGBLSCAL}
and SciRepEval \cite{Singh2022SciRepEvalAM} that measure performance on a single graph.  However, it is useful to see how
performance scales with graph size because of network effects (Metcalfe's Law \cite{metcalfe2013metcalfe}):
nodes, edges and paths scale with $n$, $n^2$ and $2^n$, respectively.
Network effects  are important
for academic search because the literature is large and growing exponentially, doubling between 9 and 19 years\footnote{\url{https://blogs.nature.com/news/2014/05/global-scientific-output-doubles-every-nine-years.html}} 
\cite{bornmann2021growth}.
Most benchmarks are smaller than the crossover point in \autoref{fig:prone-specter-crossover2}  and \autoref{tab:crossover}.

\begin{wrapfigure}{r}{0.22\textwidth}
    \centering
    \includegraphics[width=0.22\textwidth]{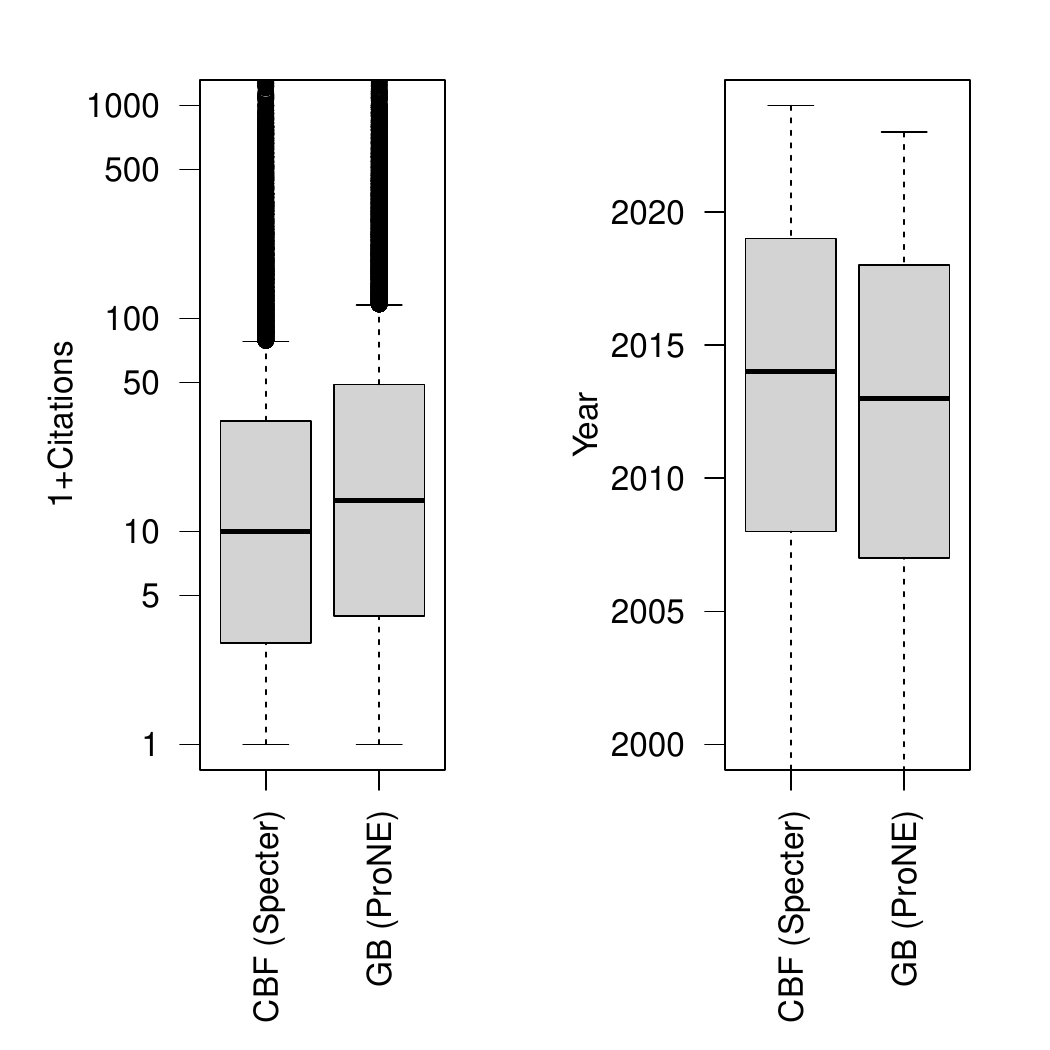}
    \caption{\label{fig:priors}  GB recommendations have more citations, but are less recent.}
\end{wrapfigure}

\subsubsection{Time Invariance}
\label{sec:time_invariance}

CBF embeddings are time invariant because abstracts do not change after a paper is published.
Authors are not allowed to change what they said after publication, but the audience
perspective is allowed to evolve over time.
It can take the literature years/decades to appreciate what is important and why.   
There are many examples of rejected papers that have subsequently received
thousands of citations such as the classic paper on Page Rank \cite{page1999pagerank}
and Broder's taxonomy of web queries \cite{broder2002taxonomy}.
From a computational perspective, time invariance is convenient.
From this
perspective, it is unfortunate that GB embeddings are not time invariant,
but the good news is that GB embeddings improve over time as papers receive more and more citations, often years
after publication.

\subsubsection{Priors}

The differences in Section \ref{sec:time_invariance}  have consequences
for priors, as shown in \autoref{fig:priors}.  GB 
is designed to return papers with more citations.  However,
because it takes time for papers to accumulate citations,
papers with more citations tend to be older.

\subsubsection{Corner Cases and Missing Values}

 \begin{figure}[b]
           \centering
    \includegraphics[width=0.9\linewidth]{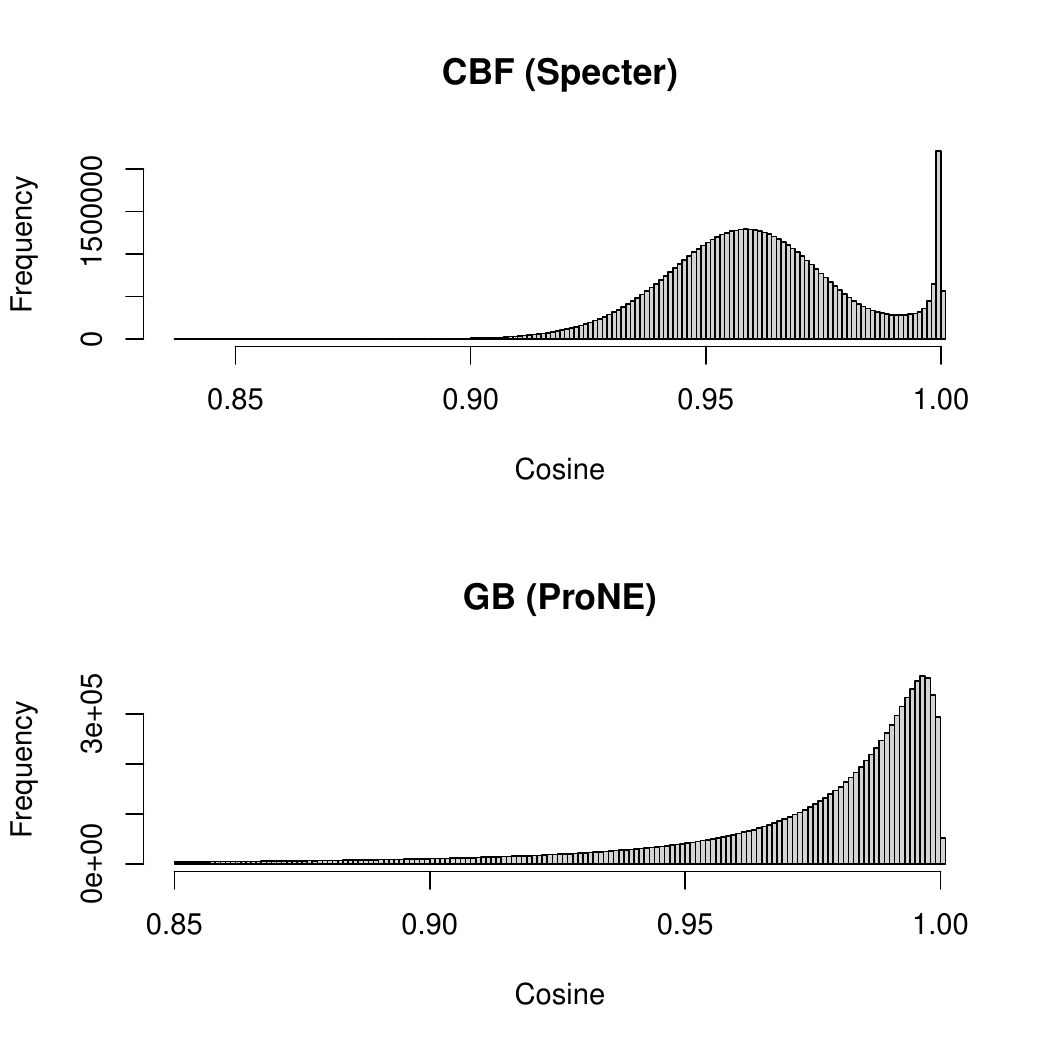}
           \caption{\label{fig:x1_cos} The spike near 1 (top) is dominated by corner cases involving abstracts.  ProNE (bottom) is not bimodal.}
         \end{figure}  

Multiple perspectives create 
opportunities for robustness. 
Corner cases can often be detected by looking for
large differences between Specter cosines and ProNE cosines.
For example, duplicates will receive large Specter cosines
because the abstracts are nearly the same, but much smaller ProNE cosines because one of the duplicates tends to have more citations than the other.

How many corner cases are there?
The spike near 1 in 
\autoref{fig:x1_cos} (top) provides a rough estimate.  This figure shows 
$cos(q,cand_1)$, the cosine between a query and the top candidate recommendation.  About 10\% of Specter vectors in S2 have a nearby neighbor with a cosine of 0.99 or better.  
ProNE is very different, with a single mode and no second spike near 1,
as shown in \autoref{fig:x1_cos} (bottom).\footnote{In addition to the number of modes, the y-axes are very different, partly because of
sample sizes (58M queries for Specter and 9M for ProNE), but
more importantly, differences in dynamic range.  ProNE cosines
have a larger dynamic range, and are centered near 0, whereas
Specter cosines can be well above 0, even for random papers.}
Spot checks of the Specter spike near 1
suggest the spike is dominated by corner cases such as duplicates, missing/incorrect
abstracts, and abstracts in languages other than English.  

The last case is perhaps the most common case.  Specter was designed for English.  When Specter is applied to Chinese, for example, the tokenizer returns many unknown tokens.
As a result, two unrelated papers in Chinese (and many other languages) can appear to be more similar to one another than they should be.
ProNE avoids these corner cases, because ProNE does not depend on abstracts, though ProNE runs into different corner cases involving papers with few (if any) links in the citation graph.

Missing values are another important corner case, as illustrated in \autoref{fig:venn}.
Imputing missing vectors goes beyond the scope of this paper, but two approximations can improve coverage:

\begin{enumerate}
  \setlength{\itemsep}{0pt}
  \setlength{\parskip}{0pt}
 \setlength{\parsep}{0pt}
    \item The centroid approximation: infer the missing vector from the average of its references
    \item The better-together approximation: infer the missing vector from the average of papers nearby in another embedding
\end{enumerate}
\noindent
The better-together approximation illustrates how CBF and GB methods complement one
another in a synergistic way.

\section{Conclusions}

\begin{table}
\small
\centering
 \begin{tabular}{p{1.8cm} | p{2.5cm} p{2.5cm}}
\textbf{Feature}& \textbf{CBF}& \textbf{GB} \\ \hline

Inputs & Titles and abstracts  & Citation graph \\  \hline

Perspective & Authors' position & Audience response\\ \hline

Technology & Deep Nets \& LLMs & Spectral Clustering \\ \hline

Discipline & Computer Science & Linear Algebra (Math) \\ \hline

Motivation & Use cases in NLP & Traffic Analysis\\ \hline 

Embedding & Specter \cite{cohan-etal-2020-specter}  & ProNE \cite{zhang2019prone} \\  \hline

Interpretation  & Similar abstracts &  
Nearby in terms  \\ 
of large cosines & & of random walks \\ \hline

Bottleneck  & Cycles & Memory \\   \hline

Hardware & GPUs & Terabytes of RAM \\ \hline

Scale & Favor smaller graphs & Favor larger graphs \\ \hline
Invariance & Abstracts are invariant after publication & 
 Citations accumulate after publication \\ \hline
 
Priors & More recent &  More impact (cites) \\ \hline

Corner Cases & Non-English abstracts & Few links in graph \\ \hline
\end{tabular}
\caption{\label{tab:features} Feature table for comparing CBF and GB methods for academic paper recommendation.}
\end{table}

Surveys group the literature into CBF, GB,
CF and hybrids methods.  This paper focused on CBF and GB because
of the availability of data from Semantic Scholar.
We summarize the main features for comparing CBF and GB methods for Academic article recommenations scenarios in Table~\ref{tab:features}.

We view the literature as a conversation between authors
and the audience, where abstracts are used by CBF methods to shed light on authors' perspectives and citations are used by GB methods to shed light on responses from the audience.  Cosines are used to estimate similarities for both CBF and GB, but large CBF cosines indicate similar abstracts whereas large GB cosines indicate similarity in terms of walks on citation graphs.  Abstracts and citations scale differently because of network effects.  Abstracts do not change after publications, unlike citations which improve with time.  GB prefers highly cited papers, but these papers tend to be older
because it takes time to accumulate citations. 
Multiple perspectives create opportunities to improve robustness and coverage.

\bibliography{refs,more_refs,refs_from_demo_paper}
\bibliographystyle{IEEEtran}

\end{document}